\documentclass{ragtime}
\usepackage{times}
\usepackage{color}  
\title[RADs  models: Polytropic equations of state and toroidal magnetic fields]%
 {On the accreting tori sequences in  Ringed Accretion Disks  models.  Polytropic equations of state and toroidal magnetic fields}

\author[D. Pugliese and 
 Z. Stuchl\'{\i}k]
 {Daniela Pugliese\at[]{1,a} 
 and Zden\v{e}k Stuchl\'{\i}k\at[]{1,b}\\
 \ins{1}Institute of Physics and Research Centre of Theoretical Physics and Astrophysics,\splitins[1]
 Faculty of Philosophy \& Science,
 Silesian University in Opava,\splitins[1]
 Bezru\v{c}ovo n\'{a}m\v{e}st\'{i} 13, CZ-74601 Opava, Czech Republic
 \\
 \ins{a}\Email{d.pugliese.physics@gmail.com}\\
 \ins{b}\Email{zdenek.stuchlik@physics.cz}}

\usepackage{amsmath,amssymb,amsthm,mathrsfs,amsfonts,dsfont}

\def\beq{\begin{equation}}
\def\eeq{\end{equation}}
\def\bea{\begin{eqnarray}}
\def\eea{\end{eqnarray}}

\newcommand{\Qa}{\mathcal{Q}}
\def\Sie{\mathcal{S}}
 \newcommand{\oo}{\mathrm{O}}

\newcommand{\pp}{\textbf{()}}

\newcommand{\il}{~}

\newcommand{\cc}{\mathrm{C}}

\def\be{\begin{equation}}
\def\ee{\end{equation}}
\def\bea{\begin{eqnarray}}
\def\eea{\end{eqnarray}}
\def\Sie{\mathcal{S}}
\graphicspath{{pug/}}

\begin{document}

\begin{abstract}
We consider ringed accretion disks (\textbf{RADs}),  representing models of   aggregates  of corotating and counterrotating toroids orbiting a central Kerr super-massive black hole (\textbf{SMBH}).
We comment on system of two-tori  governed by the   polytropic equation of state and  including a toroidal magnetic field. We  found the \textbf{RADs} leading function describing the  \textbf{RAD} inner structure and  governing  the distribution of orbiting toroidal structures and the emergence of the (hydro-mechanical) instabilities in the disk. We perform this analysis first in pure hydrodynamical models by considering one-specie perfect fluid toroids and then by considering the contribution of toroidal magnetic field.
\end{abstract}


\section{Introduction}
Active Galactic Nucleai (\textbf{AGNs}) provide a rich scenario to observe \textbf{SMBHs} interacting with
their environments. Chaotical, discontinuous accretion episodes
may leave traces in the form of matter remnants orbiting the
central attractor producing sequences of orbiting toroidal
structures with strongly different features as different rotation
orientations with respect to the Kerr \textbf{BH} where corotating and
counterrotating accretion stages can be mixed.

Motivated by these facts, ringed accretion disks (\textbf{RADs}) model
structured toroidal disks which may be formed during several
accretion regimes occurred in the lifetime of non-isolated Kerr \textbf{BHs}.
\textbf{RAD} features a system made up by several axi-symmetric matter
configurations orbiting in the equatorial plane of a single central
Kerr \textbf{SMBH}. Both corotating and counterrotating tori are possible
constituents of the \textbf{RADs}.
 This model was first introduced in \cite{Evolutive} and then detailed in \cite{ringed,open,dsystem,long,multy,proto-jet,PRD,Fi-Ringed}.

  The model strongly
binds the fluid and \textbf{BH} characteristics providing
indications on the situations where to search for
\textbf{RADs} observational evidences.
 The number of the instability points is generally
limited to n=2 and depends on the dimensionless
spin of the rotating central  attractor. The phenomenology associated with these
toroidal complex structures may be indeed very wide,  providing
a different interpretative  framework.
Obscuring and screening tori, possibly evident as traces
(screening) in x-ray spectrum emission, are also strongly
constrained.   More generally,
observational evidence is expected by the spectral features of
\textbf{AGNs} X-ray emission shape, due to X-ray obscuration and
absorption by one of the tori, providing a \textbf{RAD} fingerprint as a
radially stratified emission profile.

In Sec.\il(\ref{Sec:intro-RAD}) we introduce the model and the main definitions used throughout this article. In Sec.\il(\ref{Sec:polytro})  we focus on \textbf{RAD} with polytropic tori.
In Sec.\il(\ref{Sec:influence}) we analyze the  effects of a toroidal magnetic field in the formation of  several magnetized   accretion tori. Concluding remarks are in Sec.\il(\ref{Sec:conclu}).
Appendix\il(\ref{Sec:vin-li})  summarizes main constraints on the \textbf{RAD} structure.
\section{Ringed accretion disks}\label{Sec:intro-RAD}
Ringed accretion disk (\textbf{RAD}) is a fully general relativistic model of
axially symmetric but "knobby"
accretion disk orbiting on the  equatorial plane of a  Kerr \textbf{SMBH}. It is
  constituted by an aggregate of corotating and counter-rotating perfect fluid, one particle species, tori orbiting on the equatorial plane on one central  \textbf{BH} attractor. Because of the symmetries of the system (stationarity and axial-symmetry) the system is regulated  by the Euler equation only with a barotropic equation of state (\textbf{EoS}) $p=p(\varrho)$:
\bea
\label{E:II}&&
	T_{\mu\nu}=(p+\varrho)U_\mu U_\nu - p g_{\mu\nu},
\quad
	\frac{\nabla_\mu p}{p+\varrho}=-\nabla_\mu\ln(U_t)+\frac{\Omega \nabla_\mu \ell}{1-\Omega\ell}
\\\nonumber
&& \Omega=\frac{U^\phi}{U^t}=-\frac{g_{tt}}{g_{\phi\phi}}\ell_0=\frac{f(r)}{r^2\sin^2\theta}\ell_0,\quad \ell=-\frac{U_\phi}{U_t}.\quad
 V_{eff}(\ell)\equiv u_t\quad W\equiv\ln V_{eff}(\ell),
\eea
($(t, r, \phi, \theta)$ are  Boyer-Lindquist
 coordinates), where $V_{eff}(\ell)$ is the {torus effective potential},
$\Omega$ is the fluid relativistic angular frequency,
 $\ell$  specific angular momenta,   assumed  constant  and conserved for each \textbf{RAD} component but variable  in the \textbf{\textbf{RAD}} distribution, $U^a$ is the fluid four velocity, $T_{\mu\nu}$ is the fluid energy momentum tensor.

 We introduce the following definitions:
   we use the notation $\pp$ to indicate a configuration which can be closed, $\cc$,  or open $\oo$. Specifically,
  toroidal surfaces correspond to the   equipotential surfaces, critical points of $V_{eff}(\ell) $ as function of $r$, thus  solutions of   $W:\;  \ln(V_{eff})=\rm{c}=\rm{constant}$ or $V_{eff}=K=$constant where
{$\mathbf{C}$}--cross sections of the{ closed} surfaces (equilibrium quiescent torus);
{$\mathbf{ \cc_{\times}}$ }--cross sections of the {closed cusped}   surfaces (accreting torus);
{$\mathbf{O_{\times}}$}--cross sections of the {open cusped}   surfaces, generally associated to proto-jet configurations\cite{proto-jet,open}.
Sign $\Qa_{\pm}$ for a general quantity $\Qa$ refers to counterrotating and corotating tori respectively.
We  introduce   the concept  of
 \emph{$\ell$corotating} disks,  defined by  the condition $\ell_{(i)}\ell_{(o)}>0$, and \emph{$\ell$counterrotating}  disks defined  by the relations   $\ell_{(i)}\ell_{(o)}<0$.  The two $\ell$corotating tori  can be both corotating, $\ell a>0$, or counterrotating,  $\ell a<0$, with respect to the central attractor spin $a>0$.
We use short notation  $\pp_i<\pp_o$  and  $\pp_o>\pp_i$ for the inner and outer configurations of a \textbf{RAD} couple.

An essential part of the \textbf{\textbf{RAD}} analysis is the characterization of the boundary conditions on each torus in the agglomerate and  of the \textbf{RAD} disk inner structure. The model is constructed  investigating the
 function representing the  angular momentum distribution inside the disk (which is not constant), which  sets the toroids  location (and equilibrium) in the agglomerate and it coincides, in the hydrodynamical \textbf{RAD} model of perfect fluids,  with the distribution of specific angular momentum of the fluid in each agglomerate toroid. This function can be written as
\bea&&\label{Eq:laud}
\textbf{Leading (HD) RAD function}\quad \pm\ell^{\mp}=\pm\left.\frac{a^3+a r (3 r-4M)\mp\sqrt{r^3 \Delta^2}}{a^2-(r-2M)^2 r}\right|_{r^{\ast}},
\\
&&\nonumber \Delta\equiv r^2-2 M r+a^2,
\eea
($M$ is the central \textbf{BH} mass). Each point $r>r_{mso}$ (marginally stable orbit) on curve $\ell^{\mp}$ fixes  the center (points of maximum density inside the torus) of  the toroidal \textbf{\textbf{RAD}}  component, $r<r_{mso}$ sets    possible instabilities points of the toroids, more details can be found in \cite{ringed,dsystem}.
Because of the importance of this function in defining the inner structure of the \textbf{RAD} this is called Leading  \textbf{RAD} function. We shall see in Sec.\il(\ref{Sec:influence}) that, by changing  the  energy momentum  tensor including for example a toroidal magnetic field, it will be convenient to change the leading function $\ell^{\pm}$  adopted in the Hydrodynamical (HD) case  to a different function,
 obtained through the study of the magnetic field in the \textbf{RAD} and  able to represent and regulate  the tori distribution. In Sec.\il(\ref{Sec:vin-li}) we include a summary of the main constraints on the \textbf{RAD} inner structure--\cite{ringed,dsystem}
\subsection{Polytropic tori}\label{Sec:polytro}
We conclude
 this section,
 considering   \textbf{RAD} tori   with  polytropic fluids:  $p=\kappa \varrho^{1+1/n}$.  We develop some general considerations on the \textbf{EoS} and the  polytropic  \textbf{RAD} tori governed  by the \textbf{EoS}: $p=\kappa \varrho^{\gamma}$, where  $\kappa>0$ is a constant   and $\gamma=1+1/n$ is the polytropic index, in \cite{PuMonBe12}. Details on this analysis can be found in  \cite{PRD} we also refer to this analysis for a comment on the tori  energetics of various \textbf{RAD} configurations, and significance in the case of polytropic tori.
It has been shown in \cite{PuMonBe12,PRD} that for the Schwarzschild geometry $(a=0)$ there is a specific classification of eligible geometric polytropics  and a specific class of polytropics is characterized by a discrete  range of values for the index $\gamma$.
Therefore, we can   propose a general classification for the tori  ($\cc, \cc_{\times}$), as for proto-jets $\oo_{\times}$, assuming a particular representation of the density function. We can write  the density $\varrho$ as function  $\gamma$.
However, we concentrate our attention   on the \textbf{RAD} components  $\cc$ and $\cc_{\times}$ for which  $K<1$, there is :
\bea\label{Eq:pe-t-r}
&&
\varrho_{\gamma}\equiv
\kappa^{1/(\gamma-1)}\bar{\varrho}_{\gamma}\quad\mbox{and}\quad
\bar{\varrho}_{\gamma}\equiv\left[\frac{1}{\kappa}\left(V_{eff}^{-\frac{\gamma -1}{\gamma
}}-1\right)\right]^{\frac{1}{(\gamma-1)}}\;\mbox{for}
\\
&&\nonumber\gamma\neq1 \quad
\mbox{with}\quad
\varrho_{\gamma}\equiv \mathds{C}^{1/(-1+\gamma )},\quad  \mathds{C}\equiv (V_{eff}^{-2})^{\frac{\gamma-1}{2 \gamma }}-1.
\eea
 (note $\mathds{C}$ is actually a function of  $K\in ]K_{\min},K_{\max}]$, while  $K_{\max}<1$, regulates whether the torus is quiescent or in accretion).
 The pressure $p$, associated to the solution in Eq.\il(\ref{Eq:pe-t-r}),  depends on $k^{\frac{1}{1-\gamma }}$. It decreases with $\kappa$ more slowly  then $\varrho$.

We consider the case  $K<1$ with the condition  $\varrho>0$, verified, according to Eq.\il(\ref{Eq:pe-t-r}), for  $\gamma>1$.
Integration of the  $\varrho$ density function in the polytropic case where
$\gamma=4/3$ is shown in Figs\il(\ref{Fig:MediaRT}). The situation for  different indices, and particularly  $\gamma=5/3$, is also shown, integration of density profiles have been specified particularly for the couple  $\cc^-_{\times}<\cc^+_{\times}$.
Note that  we can then directly impose several  constraints  for the density function. Some simple examples,  including special (composite) density profiles are, for example, the case
 $\varrho_{[+]}^{-}= \varrho_\gamma^i-\varrho_\gamma^o=\varrho_\Phi=$constant,  where \[K_i= \left(\left[\varrho_\Phi-\epsilon \left({K_o}^{-\frac{\gamma -1}{\gamma }}-1\right)^{\frac{1}{\gamma -1}}\right]^{\gamma -1}+1\right)^{-\frac{\gamma }{\gamma -1}}.\]
 In  Figs\il(\ref{Fig:MediaRT})  we show  also the profiles  $\varrho_\Phi$ for  $\epsilon=-1$. Importantly, we note that these relations should be  generally seen as constraints on \emph{independent} solutions for each \textbf{RAD} components. Toroidal    configurations emerging from these constraints ($\varrho_{[+]}^{-}$) as in Fig\il(\ref{Fig:MediaRT})    are by no means directly matched with solutions for two different \textbf{RAD} components coupled only through the  background. The   \textbf{RADs} effective potential (a potential describing the entire macrostructure as introduced in \cite{ringed}) may be derived from  composite energy-momentum tensors made by  collections of  the  fluid tensors  decomposed in each fluid adapted frame. This holds for not colliding tori.  They  will be naturally coupled  through the  unique  background metric tensor $g_{\mu \nu}$ and proper  boundary conditions imposed on the fluid density and pressure. The boundary conditions by the  step-functions cuts $H(\theta)$, defining the \textbf{RAD} in the two forms of the \textbf{RAD} potential functions,     will be included in the energy momentum tensor. Nevertheless, clearly  the projection after $3 + 1$ decomposition, defining the  3D hyperplane $h^{(i)}_{ij}$, has to be done according to the orthogonality condition defining fluids  field velocity vectors  $\mathbf{u}^{(i)}$, respectively,  for the $(i)-$torus. These solutions create special tori surfaces from the condition on the  constant pressure. Moreover,  within the limits  considered  before, these constraints  can found application also in the collision analysis, to infer the final states (es. final merger tori) from these constraints. ({Other notable cases might be  founded by the constraints
 $\varrho_{[\times]}= \varrho_\gamma^i \varrho_\gamma^o=$constant,
 $\varrho_{\{\times\}}= \varrho_\gamma(K^i K^o)=$constant
 or  $\varrho_{\{+\}}^{\pm}= \varrho_\gamma(K^i\pm K^o)=$constant.). It is possible to show that not all these profiles are related to quiescent of accreting  toroids.
\begin{figure*}
\includegraphics[width=10cm]{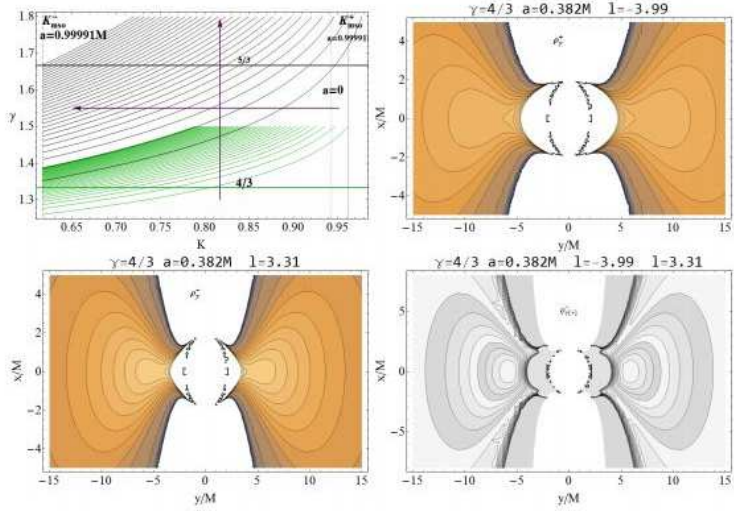}
\caption{\emph{Left panel:} Profiles of constant rationalized density function $\varrho_k$  in Eq.\il(\ref{Eq:pe-t-r}) in the plane
$\gamma-K$, $\gamma>1$ is the polytropic index, $K\in[K^{\pm}_{mso},1]$ is the $K$-parameter attached to any tori at constant $\ell$.  The values $K_{mso}^{\pm}$ for the two \textbf{SMBHs} with spin $a=0$ and $a=0.99991M$ are also plotted. Corotating (\textbf{[-]}) ad counterrotating fluids $(\mathbf{[+]})$ are  considered.   Indices $\gamma=4/3$ considered also for the analysis of \cite{PRD}  and $\gamma=5/3$ are shown.  Arrows follow the increasing values of $\varrho_{\gamma}$. The region in the  range $\gamma\in[4/3,5/3]$ has been partially thicken with
  highlighted (green-colored) $K$-constant curves.   Density profiles $\varrho_{\gamma}^{\pm}$ for corotating (\emph{bottom left panel}) and counterrotating tori  (\emph{upper  right  panel})  orbiting around a \textbf{SMBH} with spin $a=0.382 M$,  the fluids specific angular momentum is $\ell=-3.99$ and $\ell=3.31$, the polytropic index $\gamma=4/3$. $(x, y)$ are Cartesian coordinates. (\emph{Bottom right panel})  profiles of constant composite density function $\varrho_{\gamma[+]}^-$ defined in Sec.\il(\ref{Sec:polytro}).
} \label{Fig:MediaRT}
\end{figure*}
\section{Influence of toroidal magnetic field in multi-accreting tori}\label{Sec:influence}
In this section we consider \textbf{RAD} with toroidal sub-structures regulated by the presence, in the force balance equation of  a toroidal magnetic field component. We refer to the analysis of \cite{Fi-Ringed}, the toroidal magnetic field form used here is the well known Komissarov-solution \cite{Komissarov}, used in the approach \cite{epl,Fi-Ringed}, see also \cite{adamek,HK,K,zanotti}.
In this section we use mainly dimensionless units.
\subsection{Ideal GR-MHD}\label{Sec:ideal}
Before considering the model of magnetized \textbf{RAD},
it is convenient to  review  some basic notions of  ideal GR-MHD.  The  fluids energy-momentum tensor  can be written as  the composition  of the two components
\bea
\label{E:Tm}&& T^{\rm{f}}_{a b}=(\varrho +p) U_{a}
U_{b}-\epsilon\ p g_{a b}
\\\nonumber&& T^{\rm{em}}_{a
b}=-\epsilon\left (F_{a c}F^{\phantom\ c}_{b}-\frac{1}{4} F_{c d} F^{c
d} g_{ab}\right)=\frac{g_{ab}}{2}(E^2+B^2)-(E_aE_b+B_aB_b)\quad
\\\nonumber&&
-2\epsilon \mathcal{\breve{G}}_{(a}U_{b)}-\epsilon U_aU_b(E^2+B^2),\quad
\nabla_{[a}F_{b c]}=0,\quad\nabla^a F_{ab}=\epsilon J_b\quad
J^a=\varrho_c U^a +j^a,
\eea
(quantities are  measured by an observers moving
with the fluid). $\check{\mathcal{G}}_a$ denotes the Pointing vector,   $U^a U_a=\epsilon$,  ($\epsilon$ in this section is clearly a signature sign) and $
h_{ab}\equiv g_{ab}-\epsilon U_a U_b,
$  is the projection tensor,  where $\nabla_\alpha g_{\beta\gamma}=0$.
Considering the  {charge density} and  {conduction current} with the
  {Ohm's law},  there is
$
j^a=\sigma^{ab}E_b,\quad
J^a=\varrho_c U^a +\sigma E^a
$.
We consider isotropic fluids for which
$\sigma^{ab}=\sigma g^{ab}$,
$\sigma$ is the {electrical conductivity coefficient}. For ideal conductive plasma there is $\sigma\rightarrow\infty$ ($E_a=F_{ab}U^{b}=0$): the
electromagnetic field does not have a direct effect on the
conservation equation along the flow lines, or
\bea\nonumber\label{Eq:conservazione}
&&U_a\nabla^a\varrho+(p+\varrho)\nabla^aU_a- U^bF_{b}^{\phantom\ c}(\nabla^aF_{ac})=0,
\\\nonumber
&&\mbox{In the ideal MHD}
\quad
(p+\varrho)U^a\nabla_aU^c-\epsilon h^{bc}\nabla_b p-\epsilon(\nabla^aF_{ad})F^{cd}=0,\\
&&\nonumber\mbox{and}\quad
U^a\nabla_a s=\frac{1}{nT}U^b F_{b}^{\phantom\ c}\nabla^a F_{ac}.
\eea
($T$ is the temperature and $n$ is the particle number density).
In infinitely conducting plasma there is  $U^a\nabla_as=0$ the entropy per
particle is conserved  { along the flow lines of each toroids}. A particular case of
interest is when $s$ is a constant of both space and time implying $p=p(\varrho)$.
  \cite{Kroon1,Kroon2}.
\subsection{Magnetized tori}
We consider, in the magnetized case,   an infinitely conductive plasma where $F_{ab}U^a=0$, and  $F_{ab}$, $U^a B_a=0$, and   $\partial_{\phi}B^a=0$ and $B^r=B^{\theta}=0$.
The toroidal magnetic field contribution in each  \textbf{\textbf{RAD}} component  can be written by considering,
\bea&&\label{RSC}
B^{\phi }=\sqrt{\frac{2 p_B}{g_{\phi \phi }+2 \ell  g_{t\phi}+\ell ^2g_{tt}}}\quad\mbox{or alternatively}
   \\
   &&\nonumber B^{\phi }=\sqrt{{2 \mathcal{M} \omega^q}} \left(g_{t \phi }g_{t \phi }-g_{{tt}}g_{\phi \phi }\right){}^{(q-2)/2} V_{eff}(\ell)
\eea
with
\(
p_B=\mathcal{M} \left(g_{t \phi }g_{t \phi }-g_{{tt}}g_{\phi \phi }\right){}^{q-1}\omega^q
\)  the magnetic pressure,
$\omega$ is the fluid enthalpy, $q$  and $\mathcal{M}$ (magnitude) are constant;
$V_{eff}$ is a function of the metric and the angular momentum $\ell$--\cite{Komissarov,zanotti,epl,adamek,HK,K}.
The Euler  equation for the HD case  is modified by the term:
%
\bea\label{Eq:Kerr-case}
&&\partial_{\mu}\tilde{W}=\partial_{\mu}\left[\ln V_{eff}+ \mathcal{G}\right]\, \mbox{where}
\quad (a\neq0):\mathcal{G}(r,\theta)=\Sie \left(\mathcal{A} V_{eff}^2\right)^{q-1};\\ &&
\nonumber\mbox{and}\;\mathcal{A}\equiv\ell ^2 g_{tt}+2 \ell  g_{t\phi}+g_{\phi \phi },\, \Sie\equiv\frac{q \mathcal{M} \omega ^{q-1}}{q-1},
\eea
 We here concentrate on $q>1$ as, the magnetic parameter $\Sie$ is negative for $q<1$, where excretion  tori are possible\cite{excre1,excre2,excre3,excre4},   $q=1$, is a singular values for$\Sie$ ----Figs\il(\ref{Fig:splots}).
\begin{figure}
  \begin{center}
  \includegraphics[width=12cm]{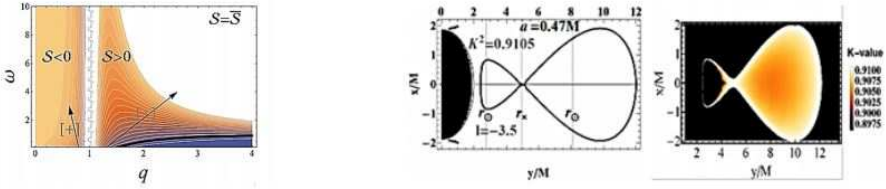}
  \caption{Left panel: Profiles of $\Sie=$constant in the panel $\omega$-q, where $\omega$ is the fluid enthalpy and $q$ is a magnetic field family parameter. Arrow directions indicate the increasing values of $\Sie$ $q=1$ is a singular value for $\Sie$. At $q<1$ ($\Sie<0$) excretion tori (density profiles in the right panel) appear. From \cite{Fi-Ringed}.}\label{Fig:splots}
  \end{center}
\end{figure}
%
In this new frame, the analysis of \textbf{RAD} structure is performed by considering the new equation
\(
\tilde{W}\equiv \mathcal{G}(r,\theta)+\ln(V_{eff})=K
\). The deformed
potential function $\widetilde{V}_{eff}^2$,
\bea\label{Eq:goood-da}
&&\widetilde{V}_{eff}^2\equiv V_{eff}^2 e^{2 \Sie \left(\mathcal{A} V_{eff}^2\right){}^{q-1}}=\\
&&\nonumber=
\frac{\left(g_{t \phi} g_{t \phi}-g_{tt} g_{\phi \phi }\right) \exp \left(2 S \left(g_{t \phi} g_{t \phi}-g_{tt} g_{\phi \phi }\right)^{q-1}\right)}{\ell ^2 g_{tt}+2 \ell  g_{t \phi}+g_{\phi \phi }}=K^2,
\eea
 for $\Sie=0$  (or $\mathcal{M}=0$) reduces to  the effective potential ${V}_{eff}^2$ for the HD case: ${V}_{eff}^2$:
$\widetilde{V}_{eff}^2= V_{eff}^2+\frac{2 \mathcal{S} \left(\mathcal{A} V_{eff}^2\right){}^q}{\mathcal{A}}+\mathbf{\mathrm{O}}\left(\Sie^2\right)$.
\bea\label{Eq:prod-gap}
\mathcal{\Sie}_n=\frac{\mathcal{M} \ln ^n(\omega ) (n+\ln (\omega )+1)}{\Gamma (n+2)}\,\mbox{for}\, n\geq0\,\mbox{ and }\, q\gtrapprox1,
\eea
where $\Gamma(x)$ is the Euler gamma function.
As for the HD case in Eq.\il(\ref{Eq:laud}),  we could  find  the \textbf{\textbf{RAD}} angular momentum distribution:
\bea\label{Eq:dilde-f}
\nonumber
&&\widetilde{\ell}^{\mp}\equiv\frac{\Delta \left(a^3+a r \left[4 \Qa (r-M) \Sie \Delta^{\Qa}+3 r-4\right]\mp\sqrt{r^3 \left[\Delta ^2+4 \Qa^2 (r-1)^2 r \Sie^2 \Delta ^{2 \Qa+1}+2 \Qa (r-1)^2 r \Sie \Delta ^{\Qa+1}\right]}\right)}{
a^4-a^2 (r-3) (r-2) r-(r-2) r \left[2 \Qa (r-1) \Sie \Delta ^{\Qa+1}+(r-2)^2 r\right]}
\\&&\label{Eq:polis-ll}
\mbox{where there is }\, \lim_{\mathcal{\Sie}\rightarrow0}\widetilde{\ell}^{\mp}=\lim_{q\rightarrow 1}\widetilde{\ell}^{\mp}=\ell^{\pm}, \quad \Qa\equiv q-1
\eea
(dimensionless units)--Figs\il(\ref{Fig:Angular}).
However  the introduction of  a toroidal magnetic field $B$, makes the study of the momentum distribution within the disk rather complicated. Instead, in \cite{Fi-Ringed}  it was adopted a function derived from the
\begin{figure}
  \includegraphics[width=11cm]{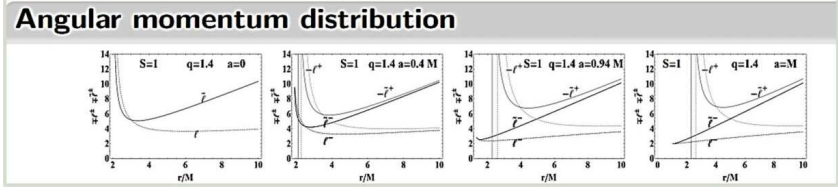}
  \caption{Magnetized \textbf{RAD}: angular momentum profiles $\tilde{\ell}$ in comparison with the HD case $\ell$, for different values of the magnetic parameters $\Sie$ and $q$ and the \textbf{BH} dimensionless spin $a/M$, from the Schwarzschild case $a=0$ to the extreme Kerr \textbf{BH} $a=M$ for corotating  (-) and counterrotating (+) fluids.  From \cite{Fi-Ringed}.}\label{Fig:Angular}
\end{figure}
$\Sie$  parameter:
\bea\label{Eq:Sie-crit}
\textbf{Leading RAD function:}\quad\mathcal{\Sie}_{crit}\equiv-\frac{\Delta^{-\Qa}}{\Qa}\frac{a^2 (a-\ell)^2+2 r^2 (a-\ell) (a-2 \ell)-4 r (a-\ell)^2-\ell^2 r^3+r^4}{2  r  (r-1)\left[r (a^2-\ell^2)+2 (a-\ell)^2+r^3\right]}
\eea
This function represents, instead of Eq.\il(\ref{Eq:dilde-f}) the new leading function for the distribution of tori in the \textbf{RAD} having  a toroidal magnetic field component (each torus is on a line  $\Sie=$constant) and able to determine \textbf{(1)} the limits on the value of the magnetic parameter for the tori formation, \textbf{(2)} the emergence of HD instability associated with the cusped configurations  $\cc_{\times}$ and $\pp_{\times}$, \textbf{(3)}  the emergence of  collision  between two tori of a \textbf{RAD} couple. \textbf{(4)} it highlights the difference between magnetized  corotating  and  counter-rotating tori with respect to the central black hole.
(This difference is  also evident from  the dependence  in  Eq.\il(\ref{Eq:Sie-crit}) from the quantities $(a\pm\ell)$.)
As demonstrated  in \cite{Fi-Ringed}, such magnetized tori  can be formed in the  \textbf{RAD} macroconfigurations  for   sufficiently small  $(q\Sie)$, the constraints described in Sec.\il(\ref{Sec:vin-li})  are essentially confirmed  for the magnetized case.
\section{Concluding remarks}\label{Sec:conclu}
The \textbf{RAD} dynamics is strongly affected bythe
the dimensionless spin of the central \textbf{BH} and the fluids relative rotation, especially in the magnetized case   considered in Sec.\il(\ref{Sec:influence}).
 More generally there is evidence of a strict correlation between \textbf{SMBH} spin,
fluid rotation and magnetic fields in \textbf{RADs} formation and evolution.
The analysis presented here  poses constraints on tori formation and emergence of \textbf{RADs} instabilities in the
phases of accretion onto the central attractor and tori collision emergence\cite{dsystem,PRD}.
Eventually the \textbf{RAD} frame investigation constraints specific classes of tori that could be observed
around some specific \textbf{SMBHs} identified by their dimensionless spin.  As sideline result we provided a full characterization of the counterrotating tori in the multi-accreting systems.
This  model is designed for an extension to a dynamic GRMHD setup.
From observational viewpoint, \textbf{AGN} Xray variability suggests connection between X-rays and the innermost regions of accretion disk. In \cite{S11etal,KS10,Schee:2008fc,Schee:2013bya}   relatively indistinct excesses
 of the relativistically broadened  emission-line components were predicted arising in a well-confined
radial distance in the accretion structure
  originating by a series of  episodic accretion events.

Another significant aspect is the possibility  to relate the \textbf{RAD}  oscillations of its  components  with  \textbf{QPOs}: radially oscillating tori of the couple could be related to
the high-frequency quasi periodic oscillations (\textbf{QPOs}).
Finally, for a discussion on the relation  between  Papaloizou-Pringle (\textbf{PP})  global incompressible modes  in the tori, the  Papaloizou-Pringle Instability
(\textbf{PPI}),
a global, hydrodynamic, non-axis-symmetric instability   and the  Magneto-Rotational Instability (MRI)  modes see  \cite{Fi-Ringed,bugli}.

As an extension of this model to more general situation  multi orbiting  configurations are also studied considering tilted of warped disks\cite{tobesubmitted}.
This possibility, rather probable as a scenario in the initial phases of tori formation, could be investigated as perturbation or deformation  of the axis-symmetric equatorial model considered here.

\ack

D.P. acknowledges partial support from the Junior GACR grant of the Czech Science Foundation No:16-03564Y.
Z. S. acknowledges  the Albert Einstein Centre for Gravitation and Astrophysics supported by grant No.
 14-37086G.

\appendix
\section{Basic HD-RAD constraints}\label{Sec:vin-li}
In this section we see some main constraints of the \textbf{RAD} models by schematically summarizing  the analysis of   \cite{dsystem,PRD}.

In general, two {quiescent tori} (not cusped tori) can exist in all Kerr  spacetimes if their specific angular momenta are properly related.
Whereas there  are only  the following four    double tori  with a critical (cusped) topology:
{\textbf{{i)}}} $\cc_{\times}^{\pm}< \cc^{\pm}$
{{\textbf{ii)}}},  $\cc_{\times}^{+}< \cc^{\pm}$,{{\textbf{iii)}}} $\cc_{\times}^{-}< \cc^{\pm}$ and
{\textbf{{iv)}}} $\cc_{\times}^{-}< \cc_{\times}^{+}$--}

Moreover: $\bullet$
for   $\ell$corotating tori  or  in the background of   a static ({Schwarzschild}) attractor    only the inner torus  can be accreting (with a cusp).
$\bullet$  In the  $\ell$counterrotating couple, an cusped   {corotating} torus {has to   be}  the inner one of the couple where the outer counterrotating torus  can be in quiescent or with a cusp. If there is  $\cc_{\times}^-$ (or   for a {static} attractor), then $\cc_{\times}^-$ is part of $\cc_{\times}^-<\cc^-$ or $\cc_{\times}^-<\pp^+$, doubled system.

%
%
 Therefore, summarizing the situation for  corotating  and counterrotating \textbf{RAD} components,  in particular   there is:
 $\bullet$
 A  corotating torus can be the outer  of a couple with an   inner counterrotating cusped surface. The outer torus of this couple may be corotating (quiescent), or counterrotating cusped  or in quiescence.    Both  the inner corotating  and the outer counterrotating  torus of the couple  can have a cusp.
$\bullet$
A counterrotating  torus can  reach the (HD) instability as the inner configuration of an $\ell$corotating or $\ell$counterrtoating couple or, viceversa, the  outer torus of an $\ell$counterrtoating  couple.
If  the {cusped} torus is  $\cc_{\times}^+$, it follows that   there is {no} inner  counterrotating torus, but there can be   $\cc_{\times}^+<\cc^{\pm}$ or $\pp^-<\cc_{\times}^+$.

\def\prc{Phys. Rev. C }
\def\pre{Phys. Rev. E }
\def\prd{Phys. Rev. D }
\def\jcap{Journal of Cosmology and Astroparticle Physics }
\def\apss{Astrophysics and Space Science }
\def\mnras{Monthly Notices of the Royal Astronomical Society }
\def\apj{The Astrophysical Journal }
\def\aap{Astronomy and Astrophysics }
\def\actaa{Acta Astronomica }
\def\pasj{Publications of the Astronomical Society of Japan }
\def\apjl{Astrophysical Journal Letters }
\def\pasa{Publications Astronomical Society of Australia }
\def\nat{Nature }
\def\physrep{Physics Reports }
\def\araa{Annual Review of Astronomy and Astrophysics}
\def\apjs{The Astrophysical Journal Supplement}
\def\na{New Astronomy}

\def\mdash{---}

\bibliography{pug-RAD}

\begin{thebibliography}{28}
\expandafter\ifx\csname natexlab\endcsname\relax\def\natexlab#1{#1}\fi
\expandafter\ifx\csname url\endcsname\relax
  \def\url#1{\texttt{#1}}\fi
\expandafter\ifx\csname urlprefix\endcsname\relax\def\urlprefix{URL }\fi
\providecommand{\selectlanguage}[1]{\relax}
\providecommand{\eprint}[2][]{\url{#2}}

\bibitem[{exc()}]{excre3}
 (????).

\bibitem[{Sch()}]{Schee:2013bya}
 (????).

\bibitem[{Ad{\'{a}}mek and Stuchl{\'{\i}}k(2013)}]{adamek}
Ad{\'{a}}mek, K. and Stuchl{\'{\i}}k, Z. (2013), Magnetized tori in the field
  of kerr superspinars, \emph{Classical and Quantum Gravity}, \textbf{30}(20),
  p. 205007.

\bibitem[{Bugli et~al.(2018)Bugli, Guilet, Müller, Del~Zanna, Bucciantini and
  Montero}]{bugli}
Bugli, M., Guilet, J., Müller, E., Del~Zanna, L., Bucciantini, N. and Montero,
  P.~J. (2018), {Papaloizou-Pringle instability suppression by the
  magnetorotational instability in relativistic accretion discs}, \emph{Mon.
  Not. Roy. Astron. Soc.}, \textbf{475}, p. 108, \eprint{1707.01860}.

\bibitem[{Hamersky and Karas(2013)}]{HK}
Hamersky, J. and Karas, V. (2013), {Effect of the toroidal magnetic field on
  the runaway instability of relativistic tori}, \emph{Astron. Astrophys.},
  \textbf{555}, p. A32, \eprint{1305.6515}.

\bibitem[{Karas et~al.(2014)Karas, Kopácek, Kunneriath and Hamerský}]{K}
Karas, V., Kopácek, O., Kunneriath, D. and Hamerský, J. (2014), {Oblique
  magnetic fields and the role of frame dragging near rotating black hole},
  \emph{Acta Polytech.}, \textbf{54}(6), pp. 398--413, \eprint{1408.2452}.

\bibitem[{{Karas} and {Sochora}(2010)}]{KS10}
{Karas}, V. and {Sochora}, V. (2010), {Extremal Energy Shifts of Radiation from
  a Ring Near a Rotating Black Hole}, \emph{\apj}, \textbf{725}, pp.
  1507--1515, \eprint{1010.5785}.

\bibitem[{{Komissarov}(2006)}]{Komissarov}
{Komissarov}, S.~S. (2006), {Magnetized tori around Kerr black holes: analytic
  solutions with a toroidal magnetic field}, \emph{\mnras}, \textbf{368}, pp.
  993--1000, \eprint{astro-ph/0601678}.

\bibitem[{Pugliese and Kroon(2012)}]{Kroon2}
Pugliese, D. and Kroon, J. A.~V. (2012), {On the evolution equations for ideal
  magnetohydrodynamics in curved spacetime}, \emph{Gen. Rel. Grav.},
  \textbf{44}, pp. 2785--2810, \eprint{1112.1525}.

\bibitem[{Pugliese and Montani(2013)}]{epl}
Pugliese, D. and Montani, G. (2013), {Squeezing of toroidal accretion disks},
  \emph{EPL}, \textbf{101}(1), p. 19001, \eprint{1301.1557}.

\bibitem[{Pugliese and Montani(2015)}]{Evolutive}
Pugliese, D. and Montani, G. (2015), {Relativistic thick accretion disks:
  morphology and evolutionary parameters}, \emph{Phys. Rev.}, \textbf{D91}(8),
  p. 083011, \eprint{1412.2100}.

\bibitem[{Pugliese and Montani(2018)}]{Fi-Ringed}
Pugliese, D. and Montani, G. (2018), {Influence of toroidal magnetic field in
  multiaccreting tori}, \emph{Mon. Not. Roy. Astron. Soc.}, \textbf{476}(4),
  pp. 4346--4361, \eprint{1802.07505}.

\bibitem[{Pugliese et~al.(2013)Pugliese, Montani and Bernardini}]{PuMonBe12}
Pugliese, D., Montani, G. and Bernardini, M.~G. (2013), {On the Polish doughnut
  accretion disk via the effective potential approach}, \emph{Mon. Not. Roy.
  Astron. Soc.}, \textbf{428}(2), pp. 952--982, \eprint{1206.4009}.

\bibitem[{{Pugliese} and {Stuchl{\'{\i}}k}(2017)}]{tobesubmitted}
{Pugliese}, D. and {Stuchl{\'{\i}}k}, Z. (2017), {to be submitted}, \emph{{}}.

\bibitem[{Pugliese and Stuchlik(2018{\natexlab{a}})}]{proto-jet}
Pugliese, D. and Stuchlik, Z. (2018{\natexlab{a}}), {Proto-jet configurations
  in RADs orbiting a Kerr SMBH: symmetries and limiting surfaces}, \emph{Class.
  Quant. Grav.}, \textbf{35}(10), p. 105005, \eprint{1803.09958}.

\bibitem[{Pugliese and Stuchlik(2018{\natexlab{b}})}]{multy}
Pugliese, D. and Stuchlik, Z. (2018{\natexlab{b}}), {Relating Kerr SMBHs in
  active galactic nuclei to RADs configurations}, \emph{Class. Quant. Grav.},
  \textbf{35}(18).

\bibitem[{Pugliese and Stuchlik(2018{\natexlab{c}})}]{long}
Pugliese, D. and Stuchlik, Z. (2018{\natexlab{c}}), {Tori sequences as remnants
  of multiple accreting periods of Kerr SMBHs}, \emph{JHEAp}, \textbf{17}, pp.
  1--37, \eprint{1711.04530}.

\bibitem[{Pugliese and Stuchlík(2015)}]{ringed}
Pugliese, D. and Stuchlík, Z. (2015), {Ringed accretion disks: equilibrium
  configurations}, \emph{Astrophys. J. Suppl.}, \textbf{221}, p.~25,
  \eprint{1510.03669}.

\bibitem[{Pugliese and Stuchlík(2016)}]{open}
Pugliese, D. and Stuchlík, Z. (2016), {Ringed accretion disks: instabilities},
  \emph{Astrophys. J. Suppl.}, \textbf{223}(2), p.~27, \eprint{1603.00732}.

\bibitem[{Pugliese and Stuchlík(2017)}]{dsystem}
Pugliese, D. and Stuchlík, Z. (2017), {Ringed accretion disks: evolution of
  double toroidal configurations}, \emph{Astrophys. J. Suppl.},
  \textbf{229}(2), p.~40, \eprint{1704.04063}.

\bibitem[{Pugliese and Stuchlík(2019)}]{PRD}
Pugliese, D. and Stuchlík, Z. (2019), {RADs energetics and constraints on
  emerging tori collisions around super-massive Kerr Black Holes},
  \eprint{1903.05970}.

\bibitem[{Pugliese and Valiente~Kroon(2016)}]{Kroon1}
Pugliese, D. and Valiente~Kroon, J.~A. (2016), {On the locally rotationally
  symmetric Einstein–Maxwell perfect fluid}, \emph{Gen. Rel. Grav.},
  \textbf{48}(6), p.~74, \eprint{1410.1335}.

\bibitem[{Schee and Stuchlik(2009)}]{Schee:2008fc}
Schee, J. and Stuchlik, Z. (2009), {Profiles of emission lines generated by
  rings orbiting braneworld Kerr black holes}, \emph{Gen. Rel. Grav.},
  \textbf{41}, pp. 1795--1818, \eprint{0812.3017}.

\bibitem[{{Sochora} et~al.(2011){Sochora}, {Karas}, {Svoboda} and {Dov{\v
  c}iak}}]{S11etal}
{Sochora}, V., {Karas}, V., {Svoboda}, J. and {Dov{\v c}iak}, M. (2011), {Black
  hole accretion rings revealed by future X-ray spectroscopy}, \emph{\mnras},
  \textbf{418}, pp. 276--283, \eprint{1108.0545}.

\bibitem[{Stuchlik(2005)}]{excre1}
Stuchlik, Z. (2005), {Influence of the relict cosmological constant on
  accretion discs}, \emph{Mod. Phys. Lett.}, \textbf{A20}, p. 561,
  \eprint{0804.2266}.

\bibitem[{Stuchlik and Schee(2010)}]{excre4}
Stuchlik, Z. and Schee, J. (2010), {Appearance of Keplerian discs orbiting Kerr
  superspinars}, \emph{Class. Quant. Grav.}, \textbf{27}, p. 215017,
  \eprint{1101.3569}.

\bibitem[{Stuchlik et~al.(2009)Stuchlik, Slany and Kovar}]{excre2}
Stuchlik, Z., Slany, P. and Kovar, J. (2009), {Pseudo-Newtonian and general
  relativistic barotropic tori in Schwarzschild-de Sitter spacetimes},
  \emph{Class. Quant. Grav.}, \textbf{26}, p. 215013, \eprint{0910.3184}.

\bibitem[{Zanotti and Pugliese(2015)}]{zanotti}
Zanotti, O. and Pugliese, D. (2015), {Von Zeipel's theorem for a magnetized
  circular flow around a compact object}, \emph{Gen. Rel. Grav.},
  \textbf{47}(4), p.~44, \eprint{1412.6447}.

\end{thebibliography}
\end{document}